\documentclass{article}

\usepackage{graphicx} 
\usepackage[margin=2.5cm]{geometry}
\usepackage{hyperref}
\usepackage{amsmath}
\usepackage{color}
\usepackage[semicolon]{natbib}
\usepackage{authblk}
\usepackage{fancyhdr}
\begin{document}
\title{Modelling collective cell migration in a data-rich age: \\ challenges and opportunities for data-driven modelling}
\author[1,3]{Ruth E. Baker\thanks{Email: \texttt{baker@maths.ox.ac.uk}, corresponding author.}}
\author[1,3]{Rebecca M. Crossley\thanks{Email: \texttt{crossley@maths.ox.ac.uk}}}
\author[1,3]{Carles Falc\'o\thanks{Email: \texttt{falcoigandia@maths.ox.ac.uk}}}
\author[2,3]{Simon F. Martina-Perez\thanks{Email: \texttt{simon.martina-perez@medschool.ox.ac.uk}}}
\affil[1]{Mathematical Institute, University of Oxford, Andrew Wiles Building, Woodstock Road, OX2 6GG, Oxford, Oxfordshire, United Kingdom.}
\affil[2]{School of Medicine and Biomedical Sciences, University of Oxford, John Radcliffe Hospital, OX3 9DU, Oxford, Oxfordshire, United Kingdom}
\affil[3]{All authors contributed equally to this work, and authorship is alphabetical by surname.}
\date{\today}
\maketitle

\begin{abstract}
Mathematical modelling has a long history in the context of collective cell migration, with applications throughout development, disease and regenerative medicine. The aim of modelling in this context is to provide a framework in which to mathematically encode experimentally derived mechanistic hypotheses, and then to test and validate them to provide new insights and understanding. Traditionally, mathematical models have consisted of systems of partial differential equations that model the evolution of cell density over time, together with the dynamics of any associated biochemical signals or the underlying substrate. The various terms in the model are usually chosen to provide simplified, phenomenological descriptions of the underlying biology, and follow long-standing conventions in the field. However, with the recent development of a plethora of new experimental technologies that provide quantitative data on collective cell migration processes, we now have the opportunity to leverage statistical and machine learning tools to determine mathematical models directly from the data. This perspectives article aims to provide an overview of recently developed data-driven modelling approaches, outlining the main methodologies and the challenges involved in using them to interrogate real-world data relating to collective cell migration.
\end{abstract}

\vfill
\pagebreak


\section{Introduction}

Collective cell migration involves the coordinated movement of large groups of cells as a cohesive unit, and it plays a crucial role in development, disease and repair~\citep{Friedl:2009:CCM,Rorth:2011:WDG,Haeger:2015:CCM,Stehbens:2024:PIC}. Such collective behaviour is driven by complex cell-cell and cell-environment interactions, which can make it difficult to tease apart the dominant driving mechanisms. Mathematical modelling provides an ideal platform to encode different hypothesised mechanisms for collective cell migration, and to tease apart the contributions of different processes therein. Traditionally, mathematical modelling of collective cell migration has involved the researcher writing down systems of partial differential equations (PDEs) that describe how cell density, as well as any biochemical signals (e.g., chemoattractants) or substrates (e.g., extracellular matrix), evolve in time. The choice of functional forms in the equations are usually based on heuristic attempts to describe the mechanisms at play, and they generally follow standard conventions in the field. More recently, modellers have also employed individual-based descriptions of collective cell migration, with each cell modelled as a discrete entity whose location evolves according to either a differential equation or a set of rules that account for cell-cell and cell-environment interactions. These rules are, once again, chosen by the user and based on their own experiences and the practice of others in the field. There are several review articles that provide an excellent overview of previous modelling efforts in these contexts~\citep{Schumacher:2016:MAT,Buttenschoen:2020:BFC,Giniunaite:2020:MCC,Crossley:2024:MTE}.

\begin{figure}
    \centering
    \includegraphics[width=1\linewidth]{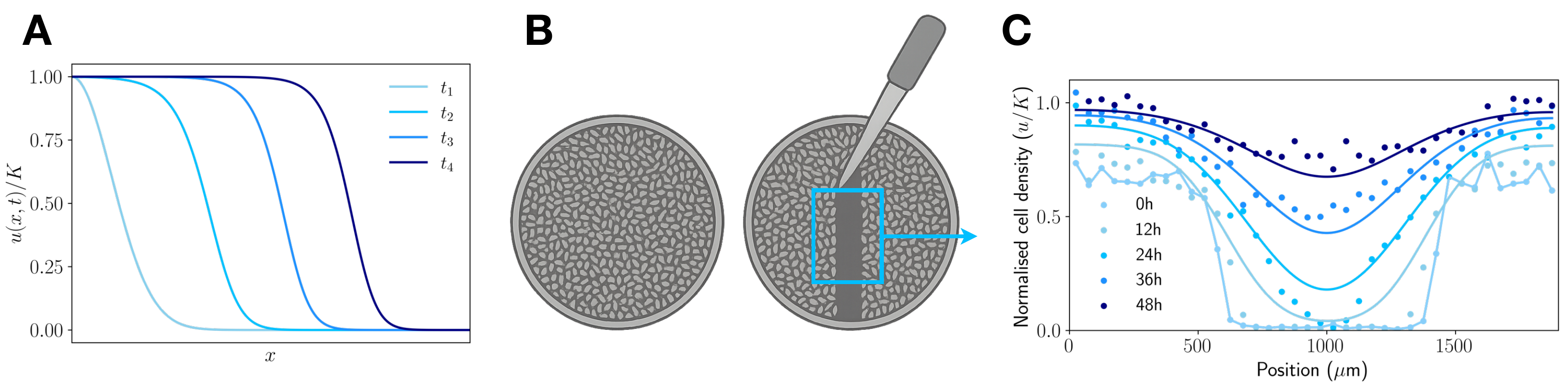}
    \caption{\textbf{Travelling wave solutions of the Fisher--KPP equation and application to cell migration in scratch assays.} (A) Numerical solutions of the Fisher--KPP equation exhibit travelling wave behaviour, with wavefronts propagating at constant speed. (B) Schematic of a scratch assay, where a cell monolayer is wounded and imaged over time to track migration into the wound area. (C) Experimental data from \cite{Jin:2016:ROS} are used to parametrise the Fisher–KPP model---$D = 1030$ $\mu$m$^2/$h, $r = 0.064$ h$^{-1}$, $K = 1.7\times 10^{-3}$ cells$/\mu$m$^2$---showing good agreement between the model and observed cell density profiles over time.}
    \label{figure:fkpp}
\end{figure}

In the simplest case, which models the evolution of cell density in one spatial dimension, $x$, over time, $t$, a PDE model can be written as
\begin{equation}
\label{equation:PDE}
\frac{\partial{u}}{\partial{t}}=\mathcal{F}\left(u;\boldsymbol{p}\right),
\end{equation}
where $\mathcal{F}$ is a differential operator that acts on $u(x,t)$ (i.e., it is a function of $x$, $t$, $u$ and its partial derivatives), and $\boldsymbol{p}$ explicitly denotes the model parameters. The equation is specified on some domain $\mathcal{D}$, and is closed by specifying relevant initial conditions, $u(x,0)=u_0(x)$ for $x\in\mathcal{D}$, as well as boundary conditions. A canonical model used for studying collective cell migration is that put forward by Fisher, as well as Kolmogorov, Petrovskii and Piskunov, and known as the Fisher--KPP equation~\citep{Murray:2003:MBI},
\begin{equation}
\label{equation:FisherKPP}
\frac{\partial{u}}{\partial{t}}=D\frac{\partial^2{u}}{\partial{x^2}}+ru\left(1-\frac{u}{K}\right),
\qquad
x\in(-\infty,\infty),
\quad
t>0,
\end{equation}
which describes the evolution of cell density due to random motility and cell proliferation, where the cell proliferation rate is a decreasing function of cell density (representing crowding effects). The parameter vector $\boldsymbol{p}=(D,r,K)$ consists of the diffusion coefficient, $D$, proliferation rate, $r$, and carrying capacity, $K$. Given suitable initial and boundary conditions, for example $u(x,0)=1$ for $x<0$ and $u(x,0)=0$ for $x\geq0$ together with $u(x,t)\to1$ as $x\to-\infty$ and $u(x,t)\to0$ as $x\to\infty$, Equation~\eqref{equation:FisherKPP} displays travelling wave solutions, which are representative of collective cell invasion---see Figure~\ref{figure:fkpp}A.


Many interdisciplinary studies on collective cell migration seek to test and validate mathematical models, and the mechanistic hypotheses encoded therein, via comparison between the model predictions and experimental observations. Historically, researchers were restricted to qualitative comparison between models and data, often using perturbation experiments to challenge model-based hypotheses. Over the past five years, however, there has been a huge transformation in the extent to which it is possible to routinely collect high-resolution spatio-temporal data on collective cell migration. As a result, many researchers now seek to calibrate their mathematical models directly to quantitative data---this allows them to estimate the model parameters and quantify the relative contributions of different mechanisms to collective migration, as well as provide a quantitative measure of the ability of their model to predict the data~\citep{Simpson:2013:EAM, Simpson:2013:QTR,Jin:2016:ROS,Nardini:2016:MKW}. Moreover, the use of Bayesian statistics approaches, for example, enables researchers to quantify the uncertainty in any parameter estimates and, with this, the uncertainty in model-based predictions~\citep{Johnston:2014:ISA,Vo:2015:QUI,Browning:2018:ABS,Stepien:2024:AAB,Falco:2023:QTG,Falco:2025:QCC}. Calibration of models to quantitative data also enables the process of model selection, in which mathematical models encoding different biological hypotheses can be quantitatively compared and selected between based on e.g., their ability to replicate the data, the extent to which model parameters can be confidently estimated from data, and model complexity~\citep{Warne:2019:UED,Liu:2024:PIA}.

Figure~\ref{figure:fkpp}B,C illustrates use of the Fisher--KPP equation in modelling collective cell migration in the context of an \textit{in vitro} wound healing assay. Here, the cells are grown to form a confluent monolayer in a dish before a portion of the cells are removed using a sharp-tipped instrument. Collective cell invasion is observed as the population recolonises the vacant scratched region. The Fisher--KPP equation can be fit to these data to provide estimates of the diffusion coefficient, $D$, proliferation rate, $r$, and carrying capacity, $K$. A comparison of the solution of the Fisher--KPP equation, given these parameter estimates, and the experimental data is shown in Figure~\ref{figure:fkpp}C---the model is able to provide accurate predictions of the cell density as the wound area is closed.

\pagebreak

An alternative approach to mathematical modelling of collective cell migration that has been enabled by the recent data revolution is that of learning the models directly from the data, herein referred to as \emph{data-driven modelling}. The aim of these methods is to infer the underlying mechanistic model directly from spatio-temporal data, by leveraging statistical and machine learning tools~\citep{Brunton2024promising,Yu2024learning}. By doing so, the hope is to eliminate the inherent bias of traditional mathematical modelling approaches, which often rely on heuristics and convention to choose the model structure and functional forms describing different biological mechanisms. Though still in their infancy, there are now a number of different methods that can be used for data-driven modelling of collective cell migration. This perspectives article aims to provide the reader with a broad overview of the field of data-driven modelling, discussing the advantages and disadvantages of the different methods in the context of learning models of collective cell motility, as well as avenues for future research. We focus on learning PDE-based models, a process that we will often refer to as \emph{equation learning}, with an emphasis on understanding what constitutes a \emph{good} model, and how to handle noise, quantify uncertainty, and develop models in a computationally efficient manner. Recent examples of methods for learning PDE-based models in the context of cell migration include~\citep{Lagergren:2020:BIN,Lagergren:2020:LPD,Martina-Perez:2021:BUQ,Simpson:2022:RAE,Nardini:2020:LEF,nardini2021learning,Nardini:2024:FAP,VandenHeuvel:2024:PCG}.

From our perspective, a \emph{good} data-driven model is one that not only fits observed data with high quantitative accuracy but also one that captures the fundamental governing dynamics (physics) of the system, rather than merely interpolating the available data, so that it can generalise beyond the regime in which it was trained to make useful quantitative predictions. 
As such, one major challenge in equation learning is achieving a balance between accurately fitting the data, and maintaining model parsimony to avoid over-fitting. A good model should also be interpretable from a mechanistic perspective, meaning that its terms correspond to meaningful physical, chemical, or biological interactions. Such interpretability allows researchers to gain deeper insights into underlying mechanisms, refine hypotheses, and potentially make novel predictions about system behaviour in untested scenarios. Moreover, a good model should have sensible mathematical properties, for example, it should preserve positivity of concentrations and densities, and satisfy relevant physical laws such conservation of mass or momentum.

Learning mathematical models directly from data inherently involves dealing with noise, measurement errors, and incomplete observations. A key question in selecting a method for equation learning is how much data a particular method requires in order to \emph{learn} a model, as well as how robust the resulting model is to variations in the data and its quality~\citep{Fajardo-Fontiveros:2023:FLT}. Here, robustness needs to be considered both in terms of the inferred model structure and in the variability of parameter estimates. For example, when it comes to noisy data, an important question to ask is whether a given method consistently gives rise to the same terms in a differential equation model, and/or the extent to which parameter estimates vary across data replicates~\citep{Martina-Perez:2021:BUQ}.

Finally, the computational complexity of data-driven methods is a critical consideration in selecting an approach, particularly when dealing with large-scale or high-dimensional data, or complicated mathematical models. One significant factor influencing the computational complexity of a method is the need to compute numerical solutions of the model at each step of the learning algorithm---this can significantly increase computational costs. Such repeated numerical integration may even render a method computationally prohibitive from a practical perspective if many thousands (or even millions) rounds of numerical integration are required. As such, it is important to analyse how computational costs scale with the amount of data and whether certain approximations or surrogate models can be used to accelerate learning while maintaining accuracy. The development of scalable algorithms that can handle large datasets efficiently will be key to making equation learning-based approaches accessible in real-world applications.

In this work, we discuss two major categories of data-driven modelling techniques in use today: symbolic regression-based approaches and neural differential equation-based approaches. In each case, we provide a brief outline of the approach, how it could potentially be used in the context of data-driven modelling of collective cell invasion (with reaction-diffusion models such the Fisher--KPP model and its relatives assumed as the underlying mathematical framework), and its advantages and disadvantages. We aim to provide only a simple description of the fundamentals of each approach here, acknowledging that many researchers have made significant contributions in improving and extending them since their original conception. Section~\ref{section:SRapproaches} covers symbolic regression-based approaches, whilst Section~\ref{section:NDEapproaches} covers those relating to neural differential equations. We conclude in Section~\ref{section:Discussion} with a brief discussion of the challenges involved in data-driven modelling and suggest avenues for possible future research.


\section{Symbolic regression-based approaches}
\label{section:SRapproaches}

One of the most influential ideas in data-driven modelling is that of \emph{symbolic regression}~\citep{bongard2007automated,schmidt2009distilling}, and one of the earliest works~\citep{Crutchfield:1987:EOM} introduced a method for reconstructing the deterministic portion of equations of motion using a function basis. This approach gained particular attention through the development of the SINDy algorithm, which learns systems of ordinary differential equations directly from data~\citep{brunton2016discovering}, and its extension to PDEs with PDE-FIND~\citep{rudy2017data}. Symbolic regression generally assumes data measured on a grid of spatial and temporal points. The approach constructs a library of candidate terms that may appear in the governing equation, evaluates them at the grid points, and performs sparse regression to identify the most relevant terms for the final model. 

More concretely, a simple application of PDE-FIND to collective cell motility would be to learn a PDE model for the evolution of cell density from noisy measurements of the cell density of the form $U^\text{data}(x_i,t_j)$, for $i=1,\ldots,n$ and $j=1,\ldots,m$. The data are assumed to be a noisy discretisation of a function $u(x,t)$ which is the solution of an unknown PDE of the form
given in Equation~\eqref{equation:PDE}. PDE-FIND then attempts to reconstruct this equation by writing the right-hand side of the PDE as a linear combination of distinct library terms,
\begin{equation}
\label{equation:PDEFIND}
\frac{\partial{u}}{\partial{t}}=\mathcal{F}^\text{\,SR}\left(u;\boldsymbol{p}\right)=\sum_{i=1}^{N_\ell}\xi_i\,\mathcal{F}^\text{\,SR}_i\left(u\right),
\end{equation}
where the $\mathcal{F}^\text{\,SR}_i$ are differential operators, each with corresponding coefficient $\xi_i$. A key assumption in PDE-FIND is that the right-hand side of the PDE contains only a few terms, and so from the large library of terms, $\left\{\mathcal{F}^\text{\,SR}_1,\ldots,\mathcal{F}^\text{\,SR}_{N_\ell}\right\}$, that make up the right-hand side of Equation~\eqref{equation:PDEFIND}, only a few should have non-zero coefficients $\xi_i$. This makes it possible to use sparse regression approaches to find these non-zero coefficients and estimate them. In this formulation, the unknown parameters $\boldsymbol{p}$ of the model are captured in the coefficients $\xi_i$---the terms $\mathcal{F}_i^\text{SR}$ may contain parameters but these cannot be estimated as part of the equation learning process.
 
The first step of the PDE-FIND pipeline is to numerically approximate both the temporal derivative on the left-hand side of Equation~\eqref{equation:PDEFIND}, and the required spatial derivatives in the library terms on the right-hand side of Equation~\eqref{equation:PDEFIND} directly from the data---this can be done, for example, using finite differences. The data and its derivatives are then combined in a matrix $F^\text{SR}(\boldsymbol{U}^\text{data})$, where each column of $F^\text{SR}$ contains all of the values of a particular candidate function across the entire $n\times m$ grid and $\boldsymbol{U}^\text{data}$ is a matrix consisting of the complete dataset. The result of approximating each of the terms in Equation~\eqref{equation:PDEFIND} is a linear matrix equation representing the PDE evaluated at the data points,
\begin{equation}
\boldsymbol{U}^\text{data}_t = F^\text{SR}(\boldsymbol{U}^\text{data})\boldsymbol{\xi},
\end{equation}
where $\boldsymbol{\xi}=[\xi_1,\ldots,\xi_{N_\ell}]^T$. Each row in this matrix equation represents the governing dynamics behind the data at one point in time and space\footnote{Generally, $F^\text{SR}$ is overspecified which means that the dynamics can be represented as linear combinations of the columns of $F^\text{SR}$.}. PDE-FIND then proceeds to learn a \emph{sparse} vector $\boldsymbol{\xi}^\text{sparse}$ as a solution of Equation~\eqref{equation:PDEFIND} by optimising the fit of the data to the PDE whilst penalising for large values of the $\xi$. 

Mathematically, this process can be expressed through finding $\boldsymbol{\xi}^\text{sparse}$ that minimises the loss function
\begin{equation}
\label{equation:ridgeregression}
\mathcal{L}^\text{SR}(\boldsymbol{\xi})=\Vert \boldsymbol{U}^\text{data}_t - F^\text{SR}(\boldsymbol{U}^\text{data})\boldsymbol{\xi}\Vert_2^2 + \lambda \Vert \boldsymbol{\xi}\Vert_2^2,
\end{equation}
alternatively, 
\begin{equation}
\label{equation:ridgeregressionSolution}
\boldsymbol{\xi}^\text{sparse} = \text{argmin}_{\boldsymbol{\xi}} \, \mathcal{L}^\text{SR}(\boldsymbol{\xi}).
\end{equation}
In Equation~\eqref{equation:ridgeregression}, $\lambda>0$ is a free parameter that penalises large coefficients---this is the method of \emph{ridge regression} ~\citep{rudy2017data, kaiser2018sparse,nardini2021learning}. PDE-FIND combines ridge regression optimisation within an iterative \emph{sequential thresholding} procedure, in which a solution to Equation~\eqref{equation:ridgeregressionSolution} is found and all library terms that have coefficients smaller than some tolerance ($|\xi_i|<\text{tol}$) are removed. The process is repeated until all coefficients are larger than the threshold or some maximum number of iterations is reached. The regularising term $\Vert \boldsymbol{\xi}\Vert_2^2$ can be replaced with others, for example, $\Vert \boldsymbol{\xi}\Vert_1^2$ which corresponds to performing LASSO~\citep{Lagergren:2020:BIN,nardini2021learning} or $\Vert \boldsymbol{\xi}\Vert_0^2$ which corresponds to hard thresholding. The optimal choice is largely problem-dependent, and no method has been proven to be definitively preferred over another~\citep{kaiser2018sparse,de2020discovery}. 

\begin{figure}[tb]
    \centering
    \includegraphics[width=\linewidth]{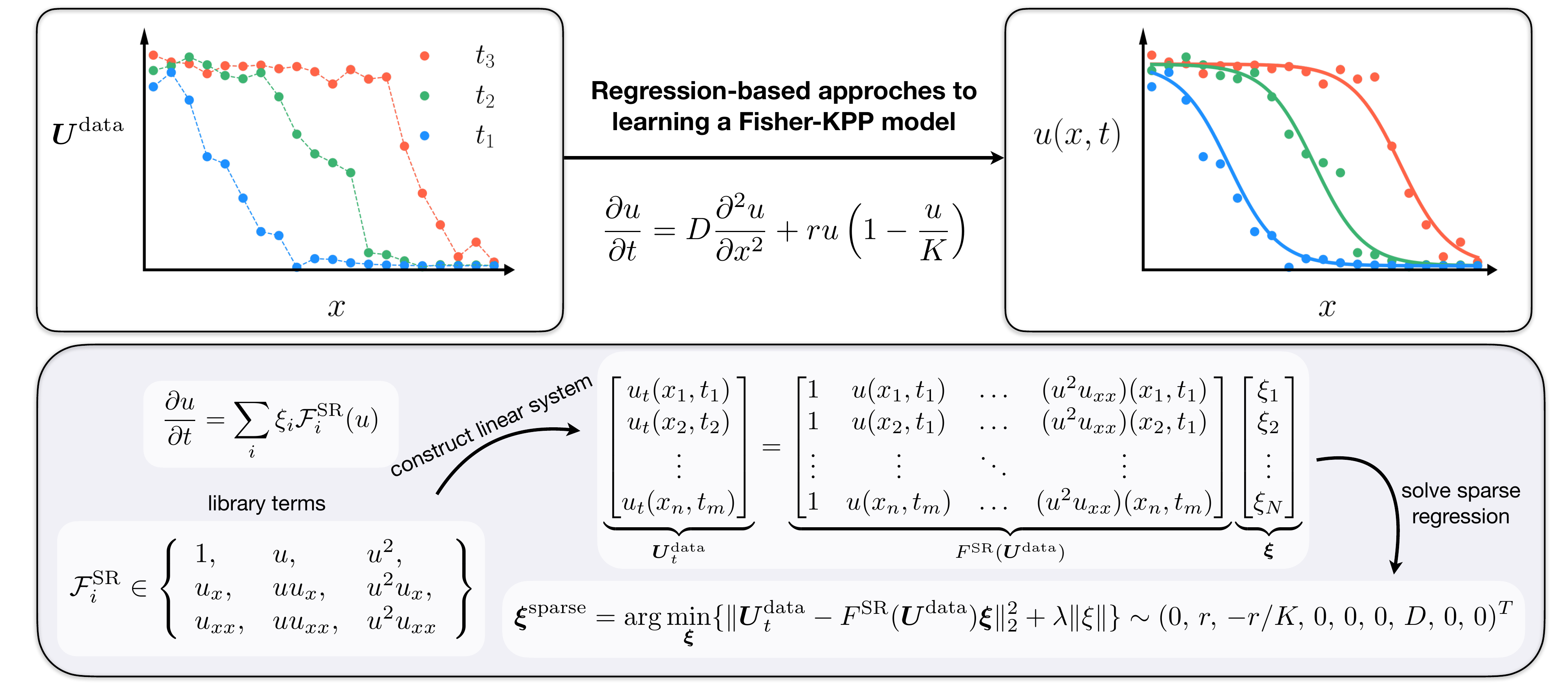}
    \caption{\textbf{Sparse regression-based approach to learning a Fisher--KPP model.} A library of candidate terms is constructed from the data, and the governing PDE is identified by solving a sparse regression problem, where only the most relevant terms (e.g., diffusion and logistic growth) are retained.}
    \label{figure:sindy}
\end{figure}

An example application of sparse regression to learn a PDE for collective cell invasion is illustrated in Figure~\ref{figure:sindy}. A library of candidate terms is constructed, for example~\citep{Martina-Perez:2021:BUQ}:
\begin{equation}
\left\{
1, u, u^2, \frac{\partial{u}}{\partial{x}},u\frac{\partial{u}}{\partial{x}},u^2\frac{\partial{u}}{\partial{x}},\frac{\partial^2{u}}{\partial{x^2}},u\frac{\partial^2{u}}{\partial{x^2}},u^2\frac{\partial^2{u}}{\partial{x^2}}
\right\}.
\end{equation}
The parameters of the model, $\boldsymbol{p}$, appear in the library coefficients $\boldsymbol{\xi}=[\xi_1,\ldots,\xi_{N_\ell}]^T$, where each $\xi_i$ can represent for instance, a diffusion coefficient, a growth rate, a carrying capacity, etc., depending on the form of the library term. A linear matrix equation is then formed from data for the cell density, $u(x,t)$, where the time derivative $u_t$ is expressed as a linear combination of these terms. By performing sparse regression, only the most relevant terms are retained---typically recovering the diffusive and logistic growth components of the Fisher--KPP model. This approach enables interpretable model discovery directly from data.


Extensions of this symbolic regression approach have evolved rapidly, incorporating diverse methodologies to enhance performance and applicability~\citep{Brunton2024promising}. Some approaches focus on reduced coordinate representations, learning coordinate systems where nonlinear dynamics are rendered linear~\citep{kutz2016dynamic,brunton2022modern}, balancing accuracy with simplicity. On the other hand, SINDYc extends SINDy to include external inputs and feedback control~\citep{brunton2016sparse}, while SINDy-BVP generalizes the framework to boundary value problems~\citep{shea2021sindy}. AI Feynman~\citep{Udrescu2020AIFeynman} combines neural network fitting with physics-inspired techniques such as dimensional analysis, symmetry, separability, and compositionality to simplify the regression process. 

The main advantage of these methods is their interpretability and ability to produce tractable, sparse models. In addition, they are computationally efficient because they do not require solution of the differential equation itself---this is a \emph{soft constraint}. However, they are highly sensitive to noise, which can lead to incorrect or physically implausible models, for example violating conservation laws. As such, one of the most significant challenges involved in using sparse regression approaches is dealing with noisy data, and assessing the ensuing uncertainty in the learned equations. One way to mitigate the effects of noise is through the use of Bayesian inference approaches. For example,~\citet{Martina-Perez:2021:BUQ} develop Bayes-PDE-FIND to quantify the uncertainty in estimates provided by PDE-FIND, and the approach of~\citet{li2020robust} seeks to automatically distinguish noise from true dynamics using a model-free method. The challenges relating to noisy data are exacerbated when computing higher order derivatives as it is incredibly difficult to estimate them from noisy data. This issue can, in part, be addressed using \emph{weak formulations} \citep{schaeffer2017sparse,gurevich2019robust,messenger2021weak,messenger2021weak1, reinbold2020using,Reinbold2021robust}. For example, Weak-SINDy~\citep{messenger2021weak,messenger2021weak1}, which has been used by \citet{kinnunen2024inference}, for example, to infer interpretable PDE models of migration and proliferation from experimental wound healing data, avoids pointwise derivative approximations to improve model recovery from noise-free data and support more robust identification of PDEs, even in high-noise settings.

Addressing computational cost is another crucial aspect of advancing symbolic regression techniques. Using \emph{Fast Fourier Transform}-based weak formulations~\citep{schaeffer2017sparse,messenger2021weak,messenger2021weak1} significantly lowers computational demand, though symbolic regression methods inherently require large libraries to achieve good predictive power. Parallelization techniques, such as those in SINDy-PI~\citep{kaheman2019identifying}, further enhance efficiency by systematically testing library terms for sparse model selection. This approach is orders of magnitude more noise-robust than previous methods and enables the identification of complex dynamical systems.  

Ultimately, sparse regression-based approaches, although easily interpretable in terms of biological mechanism, do not lend themselves well to the kinds of data routinely collected in biology, which is generally both noisy and sparse. As such, they have rapidly made way for deep learning-based methods that focus on using neural networks to learn models directly from the data as they suffer far less from such issues. We review a number of such approaches in the following section.


\section{Neural differential equation-based approaches}
\label{section:NDEapproaches}

A neural differential equation is a type of differential equation where a neural network is used to define some, or all, of the governing dynamics of the system. Neural differential equations offer a powerful framework for combining deep learning with traditional differential equations, allowing for data-driven discovery of complex system dynamics~\citep{kidger2022neuraldifferentialequations}. In this section, we outline methods that have arisen from the basic concept of a \emph{neural partial differential equation} (NPDE), which replaces the traditional right-hand-side of the PDE given in Equation~\eqref{equation:PDE} with a neural network, to write
\begin{equation}
\label{equation:NeuralPDE}
\frac{\partial{u}}{\partial{t}}=\mathcal{N}(u;\boldsymbol{\theta}),
\end{equation}  
where $\boldsymbol{\theta}$ is a vector containing the parameters of the neural network $\mathcal{N}$. 

In the context of learning a NPDE for the evolution of cell density, $u(x,t)$, from noisy measurements of the cell density of the form $U^\text{data}(x_i,t_j)$, for $i=1,\ldots,n$ and $j=1,\ldots,m$, the parameters $\boldsymbol{\theta}$ are established through training the neural network parameters to minimise a loss function of the form
\begin{equation}
\mathcal{L}^\text{NN}(\boldsymbol{\theta})=\Vert \boldsymbol{U}^\text{data}-\boldsymbol{u}^\text{pred}(\boldsymbol{\theta})\Vert_2^2,
\end{equation}
where, again, $\boldsymbol{U}^\text{data}$ is a matrix representing the data, and $\boldsymbol{u}^\text{pred}(\boldsymbol{\theta})$ is a matrix formed from the solution $u(x,t,\boldsymbol{\theta})$ of the NDE, Equation~\eqref{equation:NeuralPDE}, evaluated at points $(x_i,t_j)$. This means that, at each stage of training of the neural network, Equation~\eqref{equation:NeuralPDE} must be solved using current values of the parameters $\boldsymbol{\theta}$ in order to evaluate the loss function. The downside of this approach is therefore the computational cost of solving Equation~\eqref{equation:NeuralPDE}, but the advantage is that the NPDE is satisfied exactly at each stage of training i.e., it is a \emph{hard constraint}. In addition, although the method is very flexible, in the sense that it does not require any prior knowledge of the underlying biological mechanisms, it does not give rise to an interpretable model, and it contains large numbers of parameters which means that it can be prone to over-fitting of the data.

We now discuss two main data-driven modelling approaches that have arisen from this concept. Firstly, universal differential equations, which constitute a hybrid approach that integrates mechanistic models with neural networks to learn missing components of a model and, secondly, physics-informed neural networks (PINNs) which use neural networks to approximate the solution of differential equations while enforcing physical constraints. 


\subsection{Universal partial differential equations}

The idea underlying universal (partial) differential equations (UPDEs) is to leverage both prior scientific knowledge and data-driven methods to improve model accuracy and interpretability~\citep{rackauckas2020universal,philipps2024universal}: it combines traditional PDE modelling approaches and NPDEs to write
\begin{equation}
\label{equation:UPDE}
\frac{\partial{u}}{\partial{t}}=\mathcal{F}(u;\boldsymbol{p})+\mathcal{N}(u;\boldsymbol{\theta}).
\end{equation}  
The term $\mathcal{F}(u;\boldsymbol{p})$ is a differential operator that incorporates \emph{known} parts of the model, with the parameters $\boldsymbol{p}$ to be estimated as part of the learning process, whilst the neural network, $\mathcal{N}(u;\boldsymbol{\theta})$, is used to infer \emph{unknown} components of the dynamics from data through learning of the parameters $\boldsymbol{\theta}$. 

Given data of the form $\boldsymbol{U}^\text{data}$, the unknown dynamics together with the parameters $\boldsymbol{p}$ are estimated by minimising a loss function of the form
\begin{equation}
\label{equation:lossUPDE}
\mathcal{L}^\text{UPDE}(\boldsymbol{p},\boldsymbol{\theta})=\Vert \boldsymbol{U}^\text{data}-\boldsymbol{u}^\text{pred}(\boldsymbol{p},\boldsymbol{\theta})\Vert_2^2,
\end{equation}
where $\boldsymbol{u}^\text{pred}(\boldsymbol{p},\boldsymbol{\theta})$ is a matrix representing the solution $u(x,t;\boldsymbol{p},\boldsymbol{\theta})$ of the UPDE, Equation~\eqref{equation:UPDE}, at points $(x_i,t_j)$. As for NPDEs, the UPDE is a \emph{hard constraint} and so Equation~\eqref{equation:UPDE} must be solved numerically at each stage of the training process. The UPDEs approach can incorporate cases where the \emph{known} model parameters, $\boldsymbol{p}$, are assumed pre-estimated (known) so that the loss function depends only on the neural network parameters $\boldsymbol{\theta}$, i.e., $\mathcal{L}^\text{UPDE}(\boldsymbol{\theta})$, or mixed cases where some parameters are assumed already known and others are not.

In Figure~\ref{figure:NNs}A, we illustrate the UPDEs approach applied to a Fisher–KPP model. The structure of the PDE is preserved, with a neural network used to represent the unknown nonlinear reaction term. Given spatiotemporal data of the cell density, the network is trained together with any unknown parameters (e.g., the diffusion coefficient) by minimising the mismatch between the predicted and observed data. This allows incorporation of mechanistic structure while leveraging the flexibility of neural networks to learn complex functional forms directly from data.

One of the key advantages of universal differential equations is their potential for reliable extrapolation, particularly when compared to purely data-driven neural network models. By embedding known physical or biological dynamics within part of the model, universal differential equations retain some grounding in established physical principles, making them more robust when predicting outside the training domain. They also offer notable flexibility, particularly for modelling complex non-linear systems such as those encountered in cellular behaviour. For example, the growing availability of high-dimensional datasets---like transcriptomics, encompassing tens of thousands of genes---presents an insurmountable challenge for traditional mechanistic modelling approaches. Universal differential equations provide a way forward by combining mechanistic modelling with neural networks that can digest and represent these large, intricate datasets. This hybrid architecture also allows for integration of multimodal data, such as genomics, proteomics, and metabolomics, within a unified framework that can learn from each data stream while remaining consistent with known dynamics.

However, these benefits come with trade-offs. Universal differential equations can suffer from poor interpretability, especially when used to integrate multiple data sources. While they may outperform existing models in predictive accuracy, the contributions of the neural network component of the model can be opaque and they have limited analytical tractability. As such, understanding how specific features drive model behaviour or mapping them back to biological mechanisms can be challenging. Moreover, universal differential equations are subject to the same issues as other neural network models, including the risk of becoming ill-posed when data are sparse. This can lead to over-fitting or produce predictions that are biologically implausible~\citep{deRooij2025Physio}. To mitigate these risks, researchers often introduce constraints—such as enforcing non-negativity~\citep{Philipps2024nonnegative} or penalizing deviations from known behaviour---but these strategies are still developing and may not fully resolve the underlying limitations.


\subsection{Physics informed neural networks}

Physics informed neural networks (PINNs) take very a different approach to using neural networks to learn differential equation models~\citep{Raissi2019physics,Pang2020}. They assume a PDE model of the form given in Equation~\eqref{equation:PDE}, i.e.,
\begin{equation}
\label{equation:PINNs}
\frac{\partial{u}}{\partial{t}}=\mathcal{F}\left(u;\boldsymbol{p}\right),
\end{equation}
where the parameters, $\boldsymbol{p}$, of the differential operator $\mathcal{F}\left(u;\boldsymbol{p}\right)$ are unknown. The PINNs approach approximates the solution of Equation~\eqref{equation:PINNs}, $u(x,t;\boldsymbol{p})$, using a neural network $u^\mathcal{N}(x,t;\boldsymbol{p},\boldsymbol{\theta})$ which has parameters $\boldsymbol{\theta}$. The \emph{underlying physics} of the problem is captured through the loss function, which incorporates terms that require the neural network prediction to fit both the observed data and the PDE:
\begin{equation}
\mathcal{L}^\text{PINN}(\boldsymbol{p},\boldsymbol{\theta})=\Vert\boldsymbol{U}^\text{data}-\boldsymbol{u}^\text{pred}(\boldsymbol{p},\boldsymbol{\theta})\Vert_2^2+\lambda\Vert\boldsymbol{u}^\text{pred}_t(\boldsymbol{p},\boldsymbol{\theta})-\boldsymbol{F}^\text{pred}(\boldsymbol{p},\boldsymbol{\theta})\Vert_2^2.
\end{equation}
This time, $\boldsymbol{u}^\text{pred}(\boldsymbol{p},\boldsymbol{\theta})$ is the matrix of neural network predictions $u^\mathcal{N}(x,t;\boldsymbol{p},\boldsymbol{\theta})$, $\boldsymbol{u}^\text{pred}_t(\boldsymbol{p},\boldsymbol{\theta})$ is a matrix comprised of the neural network partial derivative $\partial{u^\mathcal{N}}/\partial{t}$, and $\boldsymbol{F}^\text{pred}(\boldsymbol{p},\boldsymbol{\theta})$ is the matrix of the neural network PDE right-hand side $\mathcal{F}\left(u^\mathcal{N};\boldsymbol{p}\right)$, each at points $(x_i,t_j)$. 


\begin{figure}[tb]
    \centering
    \includegraphics[width=.9\linewidth]{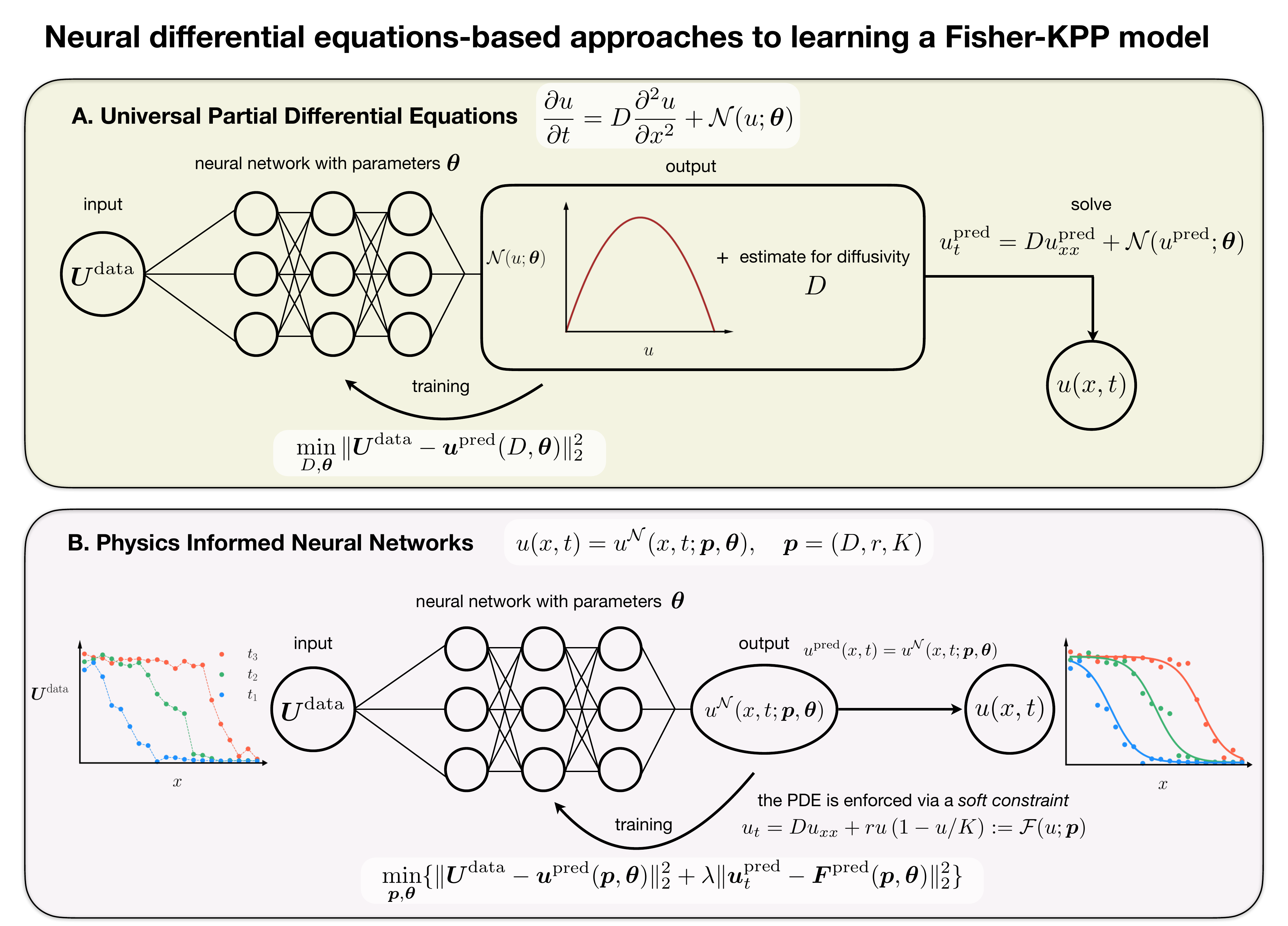}
    \caption{\textbf{Neural network-based approaches to learning a Fisher--KPP model.}
    (A) Universal differential equations approach: the neural network learns the nonlinear reaction term while the PDE structure is preserved. (B) Physics informed neural networks: the full solution is approximated by a neural network, and the PDE is enforced as a soft constraint during training.}
    \label{figure:NNs}
\end{figure}


One of the main advantages of PINNs is that the numerical solution of the PDE is not required at each step of training: automatic differentiation can be used to efficiently evaluate the second term in the loss function, $\Vert\boldsymbol{u}^\text{pred}_t(\boldsymbol{p},\boldsymbol{\theta})-\boldsymbol{F}^\text{pred}(\boldsymbol{p},\boldsymbol{\theta})\Vert_2^2$. This means that in the PINNs approach, the PDE is a \emph{soft constraint} and the hyperparameter $\lambda>0$ balances the emphasis placed on the fit of the neural network to the data (first term in the loss function) compared to the extent to which the PDE is satisfied (second term in the loss function). 

In Figure~\ref{figure:NNs}B we depict the PINNs approach for the Fisher--KPP equation, where the solution $u(x,t;\boldsymbol{p})$ is represented by a neural network. Given spatiotemporal data of the cell density, the PDE is imposed as a soft constraint during training of the neural network for the cell density through the inclusion of its residual in the loss function. This formulation enables simultaneous learning of the solution and the model parameters (the diffusion coefficient, growth rate, and carrying capacity).

A major strength of PINNs is their natural capacity to handle inverse problems: unknown parameters within a differential equation are treated as trainable variables, allowing the model to learn both dynamics and model parameters in single trainable neural network architecture. 
This flexibility is particularly useful in biological contexts, where direct estimation of model parameters is generally experimentally inaccessible. Moreover, PINNs (like UDEs) are well-suited to integrate multi-modal data---such as those arising from genomics, proteomics, and metabolomics---by learning a latent function depending on many inputs, an increasingly important feature in the era of multi-omics systems biology.

Despite its strengths, several limitations have tempered the enthusiasm surrounding PINNs. One key challenge lies in balancing the dual objectives of minimising prediction error and enforcing the differential equation: this typically involves tuning the weighting parameter ($\lambda$) that scales the relative importance of data fidelity versus physics consistency. As noted by \cite{Lagergren:2020:BIN}, poor tuning can result in models that either fit the data but violate physical constraints, or satisfy the physics but poorly capture the observed dynamics. \citet{johnston2024efficient} provide a potential solution to this approach, in the context of the method of~\citep{ramsay2007parameter,Ramsay:2017:DDA} which involves updating the weighting parameter iteratively. \citet{ramsay2007parameter} established methods for fitting models to data that avoid the use of numerical solutions of the model by employing surrogate models (for example in the form of splines) and approximately enforcing the governing differential equation. In addition, recent advances such as PINNverse~\citep{almanstotter2025pinnverse} have enhanced the reliability of PINNs for inverse problems by dynamically balancing data and equation losses during training and improving convergence and robustness even when data is noisy or the forward problem is computationally expensive.


\subsection{Combining UPDEs and PINNs}

Recent developments have sought to combine the advantages of both the UPDEs and PINNs approaches: they relax the \emph{mechanistic knowledge} requirement of the PINNs approach that the entire model structure is known in advance, letting some aspects of the right-hand side of the equation be learned using neural networks, and relax the \emph{hard constraint} requirement of the UPDEs approach that the UPDE is satisfied at each stage of the training process. 

The biologically informed neural networks (BINNs) approach, enables the user to learn the functional forms of different terms on the right-hand side of Equation~\eqref{equation:PINNs} \citep{Lagergren:2020:BIN}. For example, suppose that we anticipate that cell density evolves according to the PDE
\begin{equation*}
\frac{\partial{u}}{\partial{t}}=\frac{\partial}{\partial{x}}\left(D(u)\frac{\partial{u}}{\partial{x}}\right)+u\,G(u),
\end{equation*}
where $D(u)$ describes the density-dependent diffusion coefficient and $G(u)$ describes the density-dependent net birth rate, but that the functional forms of these functions are unknown. The BINNs approach learns these functional forms using neural networks by assuming that 
\begin{equation}
\label{equation:BINNs}
\frac{\partial{u}}{\partial{t}}=\frac{\partial}{\partial{x}}\left(D^\mathcal{\,N}(u;\boldsymbol{\theta}_\mathcal{D})\frac{\partial{u}}{\partial{x}}\right)+u\,G^\mathcal{\,N}(u;\boldsymbol{\theta}_\mathcal{G}),
\end{equation}
and approximating the solution of Equation~\eqref{equation:BINNs} using a neural network $u^\mathcal{N}(x,t;\boldsymbol{\theta}_\mathcal{D},\boldsymbol{\theta}_\mathcal{G},\boldsymbol{\theta}_\text{solution})$. The loss function is of the form
\begin{align}
\nonumber
\mathcal{L}^\text{BINN}(\boldsymbol{\theta}_\mathcal{D},\boldsymbol{\theta}_\mathcal{G},\boldsymbol{\theta}_\text{solution})&=\Vert\boldsymbol{U}^\text{data}-\boldsymbol{u}^\text{pred}(\boldsymbol{\theta}_\mathcal{D},\boldsymbol{\theta}_\mathcal{G},\boldsymbol{\theta}_\text{solution})\Vert_2^2 \\ 
& \qquad +\lambda\Vert\boldsymbol{u}^\text{pred}_t(\boldsymbol{\theta}_\mathcal{D},\boldsymbol{\theta}_\mathcal{G},\boldsymbol{\theta}_\text{solution})-\boldsymbol{F}^\text{pred}(\boldsymbol{\theta}_\mathcal{D},\boldsymbol{\theta}_\mathcal{G})\Vert_2^2,
\end{align}
where, as for UPINNs, $\boldsymbol{u}^\text{pred}$ is the matrix of neural network predictions $u^\mathcal{N}$ of the density, $u$, $\boldsymbol{u}^\text{pred}_t$ is a matrix comprised of the neural network partial derivative $\partial{u^\mathcal{N}}/\partial{t}$, and $\boldsymbol{F}^\text{pred}$ is the matrix arising from the right-hand side of the PDE, $\mathcal{F}$, each at point $(x_i,t_j)$.

The main advantage of the BINNs approach is that not all functional forms of the terms in the governing PDE need to be known \textit{a priori}. However when part of the differential operator is modelled non-parametrically some mechanistic interpretability can be lost and, as for other neural differential equation-based approaches, analytical tractability (e.g., to linear stability analysis, travelling wave analysis) is limited.

In the universal PINNs (UPINNs) approach~\citep{Podina2023}, on the other hand, the solution $u(x,t;\boldsymbol{p},\boldsymbol{\theta}_\text{model})$ of the UPDE
\begin{equation}
\label{equation:UPINNs}
\frac{\partial{u}}{\partial{t}}=\mathcal{F}(u;\boldsymbol{p})+\mathcal{N}(u;\boldsymbol{\theta}_\text{model}),
\end{equation} 
is approximated by a neural network $u^\mathcal{N}(x,t;\boldsymbol{p},\boldsymbol{\theta}_\text{model},\boldsymbol{\theta}_\text{solution})$ so that the loss function becomes
\begin{align}
\nonumber
\mathcal{L}^\text{uPINN}(\boldsymbol{p},\boldsymbol{\theta}_\text{model},\boldsymbol{\theta}_\text{solution})&=\Vert\boldsymbol{U}^\text{data}-\boldsymbol{u}^\text{pred}(\boldsymbol{p},\boldsymbol{\theta}_\text{model},\boldsymbol{\theta}_\text{solution})\Vert_2^2 \\ 
& \qquad +\lambda\Vert\boldsymbol{u}^\text{pred}_t(\boldsymbol{p},\boldsymbol{\theta}_\text{model},\boldsymbol{\theta}_\text{solution})-\boldsymbol{FN}^\text{pred}(\boldsymbol{p},\boldsymbol{\theta}_\text{model})\Vert_2^2.
\end{align}
Here, similarly to the PINNs approach, $\boldsymbol{u}^\text{pred}(\boldsymbol{p},\boldsymbol{\theta}_\text{model},\boldsymbol{\theta}_\text{solution})$ is the matrix of neural network predictions $u^\mathcal{N}$ of the density, $u$, $\boldsymbol{u}^\text{pred}_t(\boldsymbol{p},\boldsymbol{\theta}_\text{model},\boldsymbol{\theta}_\text{solution})$ is a matrix comprised of the neural network partial derivative $\partial{u^\mathcal{N}}/\partial{t}$, and $\boldsymbol{FN}^\text{pred}$ is the matrix arising from the right-hand side of the UPDE, $\mathcal{F}(u^\mathcal{N};\boldsymbol{p})+\mathcal{N}(u^\mathcal{N};\boldsymbol{\theta}_\text{model})$, each at points $(x_i,t_j)$.

The UPINNs approach combines the advantages of the UPDEs approach, in being able to learn missing physics, with that of the PINNs approach in avoiding solution of the partial differential equation at each step. However, it also suffers from some of their issues, in lacking some interpretability and analytical tractability. This approach has not yet seen application in collective cell migration, though it has been applied in other areas of the biomedical sciences~\citep{podina2024learning}.



\section{Discussion and future perspectives}
\label{section:Discussion}

In this article, we have discussed a range of approaches for learning PDE-based models of collective cell migration from experimental data, focussing on symbolic regression-based and neural differential equation-based approaches. We have described the mathematical concepts underlying each approach, and provided discussion of the advantages and disadvantages of each, as well as recent developments and applications to collective cell migration. We have grounded our examples in the Fisher--KPP equation, one of the simplest, and possibly the most widely used, model of collective cell migration. 

In addition to PDE models, significant progress has also been made in learning interaction rules in individual-based models~\citep{haupt2004practical, thede2004introduction, bottou2010large, bruckner2021learning, li2021identifiability, lu2019nonparametric, lu2020learning, lu2022data, lang2024interacting}. Data-driven approaches to infer interaction kernels in this case have allowed researchers to identify behaviours such as repulsion, adhesion, and alignment in collective migration systems. These kernels have been learned using a variety of techniques including regularized least squares~\citep{lu2019nonparametric}, nonparametric regression~\citep{lu2020learning}, approximation theory~\citep{miller2023learning}, reproducing kernel Hilbert spaces (RKHS)-based methods~\citep{lang2022learning}, and machine learning frameworks such as neural networks and integral operator learning~\citep{lu2024nonparametric}. Recent work has also explored identifiability and inference in stochastic interacting particle systems using Bayesian frameworks~\citep{lang2023identifiability,tang2024identifiability}. There are a number of comprehensive reviews written on this subject, such as those by~\citet{lu2021learning} and~\citet{lang2022learning}.

Despite the successes in applying data-driven methods to learning mathematical models of biological systems, a number of technical and conceptual challenges remain. First, identifiability and uncertainty quantification remain critical bottlenecks in data-driven modelling. Questions about the uniqueness and robustness of the inferred model structure and parameters must be addressed, particularly in ill-posed inverse problems or when observational data is sparse or noisy~\citep{li2021identifiability,tang2024identifiability}. Similarly, learning model components that are often overlooked---such as boundary conditions, non-local terms, or spatially dependent forcing functions---poses challenges that are only beginning to be explored \citep{long2019pdenet,sun2020surrogate,tartakovsky2020physics,bunge2022nonlocal,yuan2022pinn}.

Another area that requires active development concerns the implementation choices that influence the learning process. Sparse regression methods, for instance, rely on careful selection of basis functions and regularization parameters~\citep{messenger2022learning}, while neural network approaches for learning PDEs demand choices about the architecture, loss formulation, and inductive biases~\citep{Raissi2019physics}. Recent hybrid approaches, such as those combining neural and symbolic methods---for example, through physics-informed regression splines~\citep{karniadakis2021pinns}---offer promising avenues for balancing expressivity of the model with interpretability.

The increasing availability of high-resolution biological data, particularly from technologies such as omics, microscopy and barcoding, is transforming both the questions asked by experimental biologists, as well as the models developed by mathematical biologists. This data deluge offers unprecedented opportunities to formulate and validate mechanistic hypotheses. However, it also demands a shift in how mathematical modellers collaborate, design models, and interpret results. The design of experiments---ranging from the spatial and temporal resolution of data collection to the initial setup---can dramatically influence model identifiability and interpretability. Modellers can provide valuable input into experimental design, ensuring that the resulting data is optimally structured for robust inference and validation of mathematical models.  

One of the most potentially impactful opportunities from the wealth of data is the ability to learn models of increased complexity, describing for example the evolution of internal state variables of individual cells, and to interrogate and model phenotypic heterogeneity at scale. Previously, distinctions between cell subpopulations or state transitions were difficult to quantify. Now, omics data enables not only identification of distinct cellular subtypes but also inference of their roles in dynamic processes, for instance through differential gene expression or pathway activity. This growing data richness demands models that can incorporate cell heterogeneity explicitly and evolve alongside the increasing complexity of biological data~\citep{bruckner2021learning,li2021identifiability}.

In developing these models of complexity, a guiding principle should remain that models must be biologically grounded and interpretable, so that individual terms in the model can be attributed to particular biological processes, feedbacks or interactions. In addition, the learned models should be physically grounded, preserving mathematical properties such as positivity of densities, conservation of mass, and relevant boundary or symmetry conditions. Integrating these constraints explicitly---either through the model architecture or via regularization---can improve generalisability and foster biological trust in the models produced.

Looking ahead, the continuing convergence of high-resolution data and computational inference techniques opens up exciting opportunities for learning non-local models, individual cell-level behaviours, and multi-scale dynamics that reflect the true complexity of biological systems. 
However, progress will require not only methodological advances but also cultural shifts: closer and earlier collaboration with experimentalists, shared design of data collection strategies, and a commitment to developing models that are both data-driven and mechanistically interpretable.
Ultimately, the future of data-driven mathematical biology lies in its ability to synthesise detailed experimental insight with sound mathematical structure. And in adopting data-driven approaches we will be better positioned to not only explain observed phenomena but also to predict and control complex biological systems.


\section*{Code availability}

Code to produce Figure~\ref{figure:fkpp} is available on \href{https://github.com/carlesfalco/Fisher-KPP-Scratch-Assay}{Github}. Repositories for the various data-driven modelling approaches described in this chapter include:
\begin{itemize}
    \item SINDy~\citep{brunton2016discovering} -- \href{https://github.com/dynamicslab/pysindy}{PySINDy};
    \item PDE-FIND~\citep{rudy2017data} -- \href{https://github.com/snagcliffs/PDE-FIND}{PDEFIND};
    \item Universal differential equations~\citep{rackauckas2020universal} -- \href{https://github.com/ChrisRackauckas/universal_differential_equations}{UPDEs};
    \item BINNs~\citep{Lagergren:2020:BIN} -- \href{https://github.com/jlager/BINNs}{BINNS}.
\end{itemize}


\section*{Acknowledgements}

REB is supported by a grant from the Simons Foundation (MP-SIP-00001828). RMC is supported by the Engineering and Physical Sciences Research Council (EP/T517811/1). CF acknowledges support via a fellowship from ``la Caixa'' Foundation (ID 100010434) with code LCF/BQ/EU21/11890128. SMP would like to thank the Foulkes Foundation for funding. For the purpose of open access, the authors have applied a CC BY public copyright licence to any author accepted manuscript arising from this submission.


\linespread{1.0}
\bibliography{references}
\bibliographystyle{abbrvnat}


\end{document}